\documentclass[noeprint,twocolumn,showpacs,amsmath,amssymb,sort&compress,floatfix,superscriptaddress,prl,aps]{revtex4-2}

\usepackage{graphicx}
\usepackage{dcolumn}
\usepackage[normalem]{ulem}
\usepackage{bm}
\usepackage[hypertexnames=false]{hyperref}
\usepackage[mathlines]{lineno}
\usepackage{mathtools}
\usepackage{amsmath}
\usepackage{float}
\usepackage[caption=false]{subfig}
\usepackage[english]{babel}
\usepackage{xcolor}
\usepackage{bm}
\usepackage{xparse}
\addto\captionsenglish{}
\usepackage{flushend}
\usepackage{nowidow}
\usepackage{balance}
\usepackage{placeins}
\usepackage{verbatim}
\usepackage{tabularx}

\newcommand{\fig}{Fig.~}
\newcommand{\figs}{Figs.~}
\newcommand{\eqn}{Eq.~}

\newcommand{\fd}{\mathrm{d}}

\newcommand{\abs}[1]{\left|#1\right|}
\newcommand{\avg}[1]{\left\langle#1\right\rangle}
\renewcommand{\vec}[1]{\bm{#1}}

\makeatletter
\def\maketitle{
\@author@finish
\title@column\titleblock@produce
\suppressfloats[t]}
\makeatother

\newcommand{\ueph}{SUPA, School of Physics and Astronomy, University of Edinburgh, Peter Guthrie Tait Road, Edinburgh EH9 3FD, United Kingdom}
\newcommand{\ueigc}{MRC Human Genetics Unit, Institute of Genetics and Cancer, University of Edinburgh, Western General Hospital, Crewe Road South, Edinburgh, EH4 2XU, United Kingdom}

\newcommand{\papertitle}
{Bridging-Induced Phase Separation and Loop Extrusion\\ Drive Noise in Chromatin Transcription}

\begin{document}

\title{\papertitle}

\author{Michael Chiang}
\affiliation{\ueph}
\author{Cleis Battaglia}
\affiliation{\ueph}
\author{Giada Forte}
\affiliation{\ueph}
\author{Chris A. Brackley}
\affiliation{\ueph}
\author{Nick Gilbert}
\affiliation{\ueigc}
\author{Davide Marenduzzo}
\affiliation{\ueph}

\date{\today}

\begin{abstract}
Transcriptional noise, or heterogeneity, is important in cellular development and in disease. The molecular mechanisms driving it are, however, elusive and ill-understood. Here, we use computer simulations to explore the role of 3D chromatin structure in driving transcriptional noise. We study a simple polymer model where proteins -- modeling complexes of transcription factors and polymerases -- bind multivalently to transcription units -- modeling regulatory elements such as promoters and enhancers. We also include cohesin-like factors which extrude chromatin loops that are important for the physiological folding of chromosomes. We find that transcription factor binding creates spatiotemporal patterning and a highly variable correlation time in transcriptional dynamics, providing a mechanism for intrinsic noise within a single cell. Instead, loop extrusion contributes to extrinsic noise, as the stochastic nature of this process leads to different networks of cohesin loops in different cells in our simulations. Our results could be tested with single-cell experiments and provide a pathway to understanding the principles underlying transcriptional plasticity \textit{in vivo}.
\end{abstract}

\maketitle

Transcription of DNA into RNA is a fundamental intracellular process, which determines the pattern of active and inactive genes in a cell~\cite{Alberts2022}. Transcriptional programs change in development, disease, and senescence, and hence are important to ensure the correct biological function of organisms. Experimental evidence suggests that transcription is strongly dependent on the three-dimensional (3D) structure of genes and chromatin, the DNA-protein composite polymer which provides the building block of eukaryotic chromosomes~\cite{Alberts2022,Cook2018,Ghavi2019,Kempfer2020,Winick2021,Leidescher2022}. For instance, gene activation is often linked to looping in 3D between promoters and enhancers, with most transcription events occurring in transcription foci~\cite{Papantonis2013,Cook2018}, clusters of regulatory elements and associated chromatin-binding proteins forming through a phenomenon known as bridging-induced phase separation (BIPS)~\cite{Brackley2013,Brackley2021}. Yet, the biophysical mechanisms linking between 3D structure and transcription is only partially understood. In this respect, a particularly difficult-to-explain experimental result is that genome-wide removal of cohesin proteins, which stabilize long-range chromatin loops in 3D, only has subtle effects on gene expression~\cite{Rao2017,Rebollo2022,Hsieh2022,Hafner2023}.

An important aspect of transcription is that it is highly heterogeneous within a population of phenotypically homogeneous cells~\cite{Elowitz2002,Raser2004,Kaern2005,Rodriguez2019,Eling2019}, even in the absence of any genetic or epigenetic differences. What determines this variability? On the one hand, microscopic processes driving transcription, such as RNA polymerase binding and unbinding, are intrinsically noisy due to the Brownian motion of proteins and DNA. On the other hand, cells are different due to extrinsic features, such as uneven concentration of transcription factors. Consequently, no two cells in a population have the same transcriptional output, giving rise to variable gene expression, which here we denote as ``transcriptional noise''~\cite{Eling2019}. This process is important biologically, as noise has been shown to correlate with gene evolution, with younger genes being typically more noisy~\cite{Capra2013}. Similarly, transcriptional plasticity, the ability to adapt expression patterns to different environments, also correlates with noise and increases in cancer~\cite{Flavahan2017}, with implications in cells acquiring resistance to chemotherapy. However, the molecular mechanisms of transcriptional noise remain elusive, and it is unclear to what extent these may be linked to 3D structure. 

Here, we show that two biophysical principles of chromatin organization -- BIPS~\cite{Brackley2013,Brackley2016} and loop extrusion (LE)~\cite{Alipour2012,Fudenberg2016,Takaki2021,Conte2022,Chan2023} -- combine to provide molecular mechanisms to determine transcriptional noise. First, we find that intrinsic noise depends on the emergence of non-trivial correlations in the dynamics of chromatin transcription, due to the 3D clustering of DNA regulatory elements (sequences with which proteins regulating transcription interact), which is linked to BIPS and arises due to the multivalency of chromatin-binding proteins. As BIPS drives a phase transition between a swollen and a rosette-like chromatin fiber~\cite{Brackley2016}, we relate the correlations in transcription associated with noise to structural fluctuations seen near this transition. We find that in line with this interpretation, decreasing the valence of chromatin-binding proteins reduces this noise. Second, we predict that LE, performed by structural maintenance of chromosome (SMC) complexes such as cohesin~\cite{Fudenberg2016}, is a component of extrinsic noise. This is because extrusion is a stochastic process that leads to the formation of different loops in different cells, thereby increasing the diversity of promoter-enhancer interaction networks, which in turn gives different transcriptional outputs. 

\paragraph{Model.} The effects of BIPS and LE on transcription are investigated through a coarse-grained polymer model~\cite{Barbieri2012,Brackley2013,Cook2018,Brackley2020,Chiang2022}. The chromatin fiber is simulated within a cubic periodic box as a bead-and-spring chain of $N$ beads, each representing $1$ kbp of chromatin. Along the chain, a subset of beads are denoted as transcription units (TUs), modeling active \textit{cis}-regulatory elements (e.g., promoters and enhancers) that are highly accessible and have strong affinity to protein complexes such as transcription factors (TFs) and RNA polymerases [RNAPs; collectively referred to as TFs below; \fig\ref{fig:model}(a)]. TFs are simulated as multivalent chromatin-binding beads (as each represents a complex~\cite{Brackley2016,Brackley2017biophysj,Brackley2021}) and diffuse freely within the simulation box. They switch between a non-binding state and a state where they bind strongly to TU beads and weakly to other chromatin beads. These interactions enable bridging between chromatin segments, reminiscent of the promoter-enhancer looping seen in experiments~\cite{Souaid2018,Schoenfelder2019,Robson2019,Zuin2022} [\figs\ref{fig:model}(b) and (c)]. Despite there being no direct attraction between TFs, these undergo (micro)phase separation, or BIPS, as a result of positive feedback between chromatin-TF bridging and the ensuing local increase in chromatin density~\cite{Brackley2013,Cook2018}.

\begin{figure}[t!]
  \centering
  \includegraphics[width=\columnwidth]{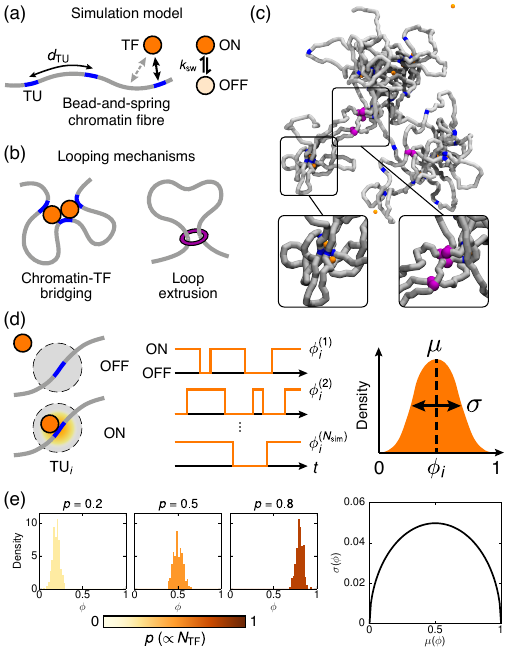}
  \caption{Polymer modeling of 3D chromatin structure and transcription. (a) Schematics of the simulation model. A chromatin fiber is modeled as a bead-and-spring chain, with certain beads denoted as TUs, separated by distance $d_{\text{TU}}$, with high affinity to TFs, which can switch between a binding (ON) and a non-binding state (OFF). (b) Chromatin loops in the simulation are driven by chromatin-TF bridges or loop extrusion (the latter are modeled by springs, mimicking cohesin complexes). (c) Simulation snapshots with TFs and cohesin loops. (d) Schematics showing how a prediction of transcriptional dynamics of each TU (blue segment) is extracted from the simulation. By measuring the fraction of time $\phi_i$ a TU is transcribed in simulation run $i$ ($i = 1,\dots,N_{\text{sim}}$; \textit{middle}), we obtain a distribution of transcriptional activities, and quantify both the average transcriptional activity $\mu$ and noise $\sigma$ of that TU (\textit{right}). (e) Distributions of transcriptional activities for TUs as predicted by sampling the baseline binomial model (\textit{left}). By varying TF number $N_{\text{TF}}$, we can vary $\mu$ and build a ``boomerang plot'', showing how $\sigma$ changes with $\mu$ (\textit{right}). }
  \label{fig:model}
\end{figure}

For simplicity, we model looping driven by SMC complexes (i.e., extruders) and halted at convergent CTCF sites by imposing $N_{\text{ex}}$ fixed loops along the chromatin fiber [see Supplemental Material (SM); \figs\ref{fig:model}(b) and (c)]. The location and size of these loops are determined based on 1D extrusion simulations~\cite{Fudenberg2016,Conte2022}, where extruders move along the chromatin fiber with velocity $v_{\text{ex}}$ and can unbind from the fiber with rate $k_{\text{off}}$. These two parameters give a lengthscale $\lambda_{\text{ex}} = v_{\text{ex}}/k_{\text{off}}$ that characterizes the typical size of a single extruded loop (without interference from others). The entire system of chromatin and TFs is simulated using Langevin dynamics, and full details of the simulation methods are given in the SM.

We predict the transcriptional activity of a TU based on measuring the fraction of time $\phi$ which TFs are associated with the TU, analogous to the binding of RNAPs to chromatin [\fig\ref{fig:model}(d)]. This quantity was shown to have a significant positive correlation with experimental data on nascent transcription~\cite{Brackley2021}. From sampling $\phi$ across many simulation runs (akin to different cells), we obtain a distribution of $\phi$ and define its mean as the mean transcriptional activity $\mu(\phi)$ of the TU and its standard deviation the transcriptional noise $\sigma(\phi)$ [\fig\ref{fig:model}(d)].

\begin{figure*}[t!]
  \centering
  \includegraphics[width=\textwidth]{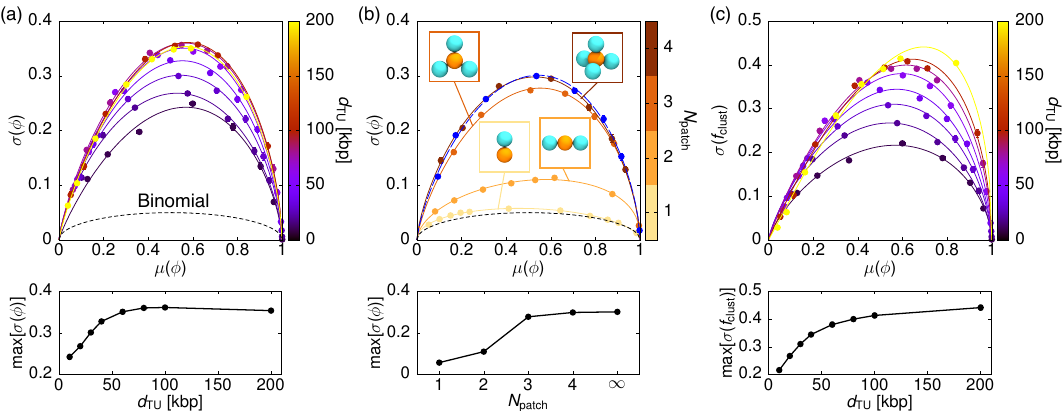}  
  \caption{Linear separation between TUs regulates transcriptional noise. (a) Boomerang plots for TU spacing $d_{\text{TU}}$ from $10$ to $200$ kbp (\textit{top}) and the maximum transcriptional noise $\sigma(\phi)$ as a function of $d_{\text{TU}}$ (\textit{bottom}). Each boomerang is obtained by varying $N_{\text{TF}}$ from $5$ to $200$, and we fit the curve $\sigma(\mu) = \frac{A_{\sigma}\mu^{\alpha}(1-\mu)^{\beta}}{\nu^{\alpha}(1-\nu)^{\beta}}$, with $\nu = \frac{\alpha}{\alpha+\beta}$, to guide the eye across different boomerangs and to extract the maximum noise [i.e., $\text{max}[\sigma(\phi)] = A_{\sigma}$]. The dashed boomerang corresponds to the binomial model, and in all cases we use $M = 101$; note that changing $M$ or total simulation time would scale all boomerang plots by the same factor. (b) Similar to (a), but showing boomerangs for patchy TFs with limited valency $N_{\text{patch}}$ (\textit{top}) and their maximal noise (\textit{bottom}), with $d_{\text{TU}} = 30$ kbp (see SM for the values of $N_{\text{TF}}$). The blue boomerang is for the case with non-patchy TFs (i.e., $N_{\text{patch}} = \infty$). Snapshots show the geometry of the chromatin-binding patches (cyan beads) on TFs. (c) Standard deviation of the fraction of time $f_{\text{clust}}$ a TU is in a cluster with other TUs as a function of activity $\mu(\phi)$ for different $d_{\text{TU}}$ (\textit{top}) and their maxima  (\textit{bottom}). Error bars representing the standard error on the mean (from averaging over TUs) are shown for both axes in all boomerangs but are smaller than the data points (same for most of subsequent boomerang plots).}
  \label{fig:TU_noise}
\end{figure*}

Within this framework, transcriptional noise $\sigma$ strongly depends on the mean expression $\mu$. This can be understood from a baseline model, where the sampling events in determining whether a TU is transcribed are independent of each other (both in time and between TUs), and the probability of transcription is $p$, which depends on the concentration of TFs. In this context, the fraction of events $\phi$ where the TU is transcribed is binomially distributed, with $\mu = p$ and $\sigma(\mu) = \sqrt{\mu(1-\mu)/M}$, where $M$ is the number of sampling events [\fig\ref{fig:model}(e)]. Given the parabolic shape of the curve $\sigma(\mu)$, we refer to this noise-mean relationship as a ``boomerang'' plot. Within the simulations, one can move along the boomerang from one end to another, for instance, by changing the concentration of TFs, as this affects the frequency with which TUs encounter TFs. In what follows, we refer to the deviation between the measured noise-mean relationship $\sigma(\mu)$ and the binomial prediction as \textit{overdispersion}.

\paragraph{Bridging-induced phase separation and intrinsic noise.} We first examine how varying the contour spacing $d_{\text{TU}}$ between uniformly-spaced TUs affects both the mean transcription $\mu$ and transcriptional noise $\sigma$, in the absence of cohesin loops arising from extrusion. \fig\ref{fig:TU_noise}(a) shows the transcription boomerangs for $d_{\text{TU}} = 10$ to $200$ kbp, and several important features are observed. First, all boomerangs show that strikingly, transcriptional noise is significantly higher in the simulation than predicted by the binomial model, indicating large overdispersion in the system. Second, the height of the boomerang increases with $d_{\text{TU}}$ up to ${\sim}100$ kbp [\fig\ref{fig:TU_noise}(a), \textit{bottom}], suggesting that TU spacing plays a fundamental role in regulating cell-to-cell variability in gene expression. 

To shed light on the mechanisms leading to the overdispersion, we plot kymographs of the TU transcription state $s_i$ (where $s_i = 1$ for transcribing and $-1$ otherwise) over time and compare them with those generated from a sequence of independent Bernoulli events [with $M = 101$; \figs\ref{sfig:trans_kymo}(a)--(d)]. Notably, the typical timescale $\tau_s$ over which a single TU remains continuously transcribed is longer than expected from Bernoulli events, as quantified by the autocorrelation $\avg{s_i(t')s_i(t'+t)}_{t',i}-\avg{s_i(t')}_{t',i}^2$ [\figs\ref{sfig:trans_kymo}(e) and (f)], and this is most apparent when there is an intermediate level of transcriptional activity. A closer look at simulation snapshots reveals that TUs that are continuously transcribed for a long time are typically in clusters with other TUs and TFs (\fig\ref{sfig:trans_clust_kymo}), suggesting that the formation of transcription factories~\cite{Papantonis2013,Cook2018} through BIPS is intimately linked to \textit{intrinsic} transcriptional noise within single cells.

\begin{figure*}[t!]
  \centering
  \includegraphics[width=\textwidth]{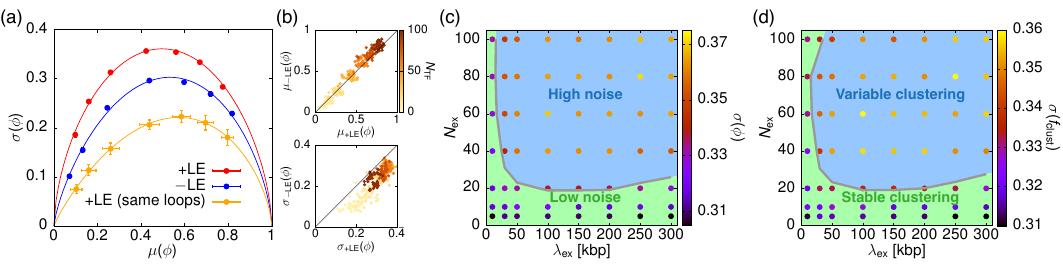}  
  \caption{Chromatin loops driven by loop-extruding factors enhance transcriptional noise. (a) Transcription boomerangs corresponding to cases with and without cohesin loops ($+$/$-$LE, respectively), and the case with loops placed at the same locations across all simulations [$+$LE (same loops)]. We vary $N_{\text{TF}}$ from $5$ to $100$, and for LE, we set the number of extruders $N_{\text{ex}} = 40$ and loop size parameter $\lambda_{\text{ex}} = 150$ kbp. (b) Scatterplots comparing the transcriptional activity $\mu$ (\textit{top}) and noise $\sigma$ (\textit{bottom}) of individual TUs between the cases with and without LE. While the data points remain close to the diagonal for $\mu$, suggesting LE has little impact on activity, they are in the lower triangle for $\sigma$, indicating that noise is higher with LE. (c) and (d) Phase diagrams showing how  $\lambda_{\text{ex}}$ and $N_{\text{ex}}$ modulate (c) transcriptional noise $\sigma(\phi)$ and (d) the fluctuations in clustering  $\sigma(f_{\text{clust}})$. Here $N_{\text{TF}} = 50$, and the gray crossover line indicates the midpoint between the minimum fluctuation in the case without extruders and the maximum fluctuation across all parameter points. $N_{\text{TU}} = 40$ and $d_{\text{TU}} = 30$ kbp for all plots.}
  \label{fig:extrusion}
\end{figure*}

To demonstrate the connection between TU clustering and noise, we modify the model such that TFs have limited valency in chromatin binding, thereby restricting their ability to bind multivalently and bridge between non-local chromatin segments to form clusters. In this version of the model, TFs are represented as patchy rigid bodies~\cite{Brackley2020JPhys}, where there are chromatin-binding beads (patches) surrounding a non-binding core [\figs\ref{fig:TU_noise}(b) and \ref{sfig:TF_patch}], and an interaction potential is selected such that each patch can only bind at most one local chromatin segment (see SM). In line with our expectations, reducing the number of patches lowers the fraction of TUs in clusters [\fig\ref{sfig:TF_patch}(e)], and the transcription boomerang falls closer to that for the binomial baseline [\fig\ref{fig:TU_noise}(b)]. This confirms that correlations due to clustering are the primary cause behind the overdispersion in the boomerang plots; interestingly, we find that just four chromatin-binding domains on TFs are sufficient to recover the case where there is no restriction on binding.

We next turn to the dependence of noise on $d_{\text{TU}}$: we find this is non-monotonic, with a weak maximum near $d_{\text{TU}} \sim 100$ kbp. This dependence can be explained by noting that changing $d_{\text{TU}}$ alters the stability of the TU clusters formed by BIPS, which are key to determining temporal correlations in the transcriptional dynamics, and hence noise. More precisely, by computing the fluctuations in the fraction of time a TU is in a cluster (i.e., two or more TUs close together) $\sigma(f_{\rm clust})$, we find that these peak at intermediate activity $\mu$ [\fig\ref{fig:TU_noise}(c)], highlighting once again that chromatin transcription is strongly linked to TU clustering in 3D through BIPS (\fig\ref{sfig:trans_clust_kymo}). For small $d_{\text{TU}}$, clusters are stable when formed, and concomitantly the fraction of time a TU is in a cluster does not fluctuate much [\fig\ref{fig:TU_noise}(c)]. Fluctuations should also disappear for $d_{\text{TU}}\to \infty$, as clusters driven by BIPS dissolve in this limit. Therefore, the temporal fluctuations in the clustering dynamics peak close to the transition when BIPS becomes effective at intermediate $d_{\text{TU}}$.

To further explore the role of 1D TU patterning on transcriptional noise, we consider chromatin fibers where TUs are randomly, rather than uniformly, positioned. For a given choice of TU positions, we find that noise is lower than that in the uniform spacing case ($d_{\text{TU}} = 30$ kbp; \fig\ref{sfig:TU_random}). We suggest that the decrease in noise is linked to an increase in cluster stability in the random fibers. This is because random TU positioning favors their clustering in 1D (through Poisson clumping), thereby decreasing fluctuations in the 3D clustering dynamics and thus noise. The dependence of noise on $d_{\text{TU}}$ provides an appealing way for cells to tune noise in different genomic regions, as $d_{\text{TU}}$ depends on the sequence and on epigenetic patterns, so that it varies across the genome and in different cell types~\cite{Brackley2021}. 

\paragraph{Cohesin loops and extrinsic noise.} We then ask how loop extrusion (LE) affects mean transcription $\mu$ and transcriptional noise $\sigma$ by incorporating loops formed by cohesin, modeled as additional springs (as detailed above). The LE dynamics means that loop positions are random between simulations, but the typical loop size and spatial correlations depend on the number of extruders $N_{\text{ex}}$ and the loop size parameter $\lambda_{\text{ex}}$. Notably, we find that transcriptional activity is affected much less by the presence of cohesin loops compared to the noise [\figs\ref{fig:extrusion}(a) and (b)]. This is in line with the experimental finding that cohesin degradation does not strongly affect gene expression levels~\cite{Rao2017,Hsieh2022}. This result points to the importance of LE in controlling the variability, rather than the mean level, of gene transcription.

To quantify how LE affects transcriptional noise, we map out a phase diagram showing the noise (at the maximum of the boomerang) as a function of $N_{\text{ex}}$ and $\lambda_{\text{ex}}$ [\fig\ref{fig:extrusion}(c)]. It shows that noise increases with $N_{\text{ex}}$, whereas the dependence on $\lambda_{\text{ex}}$ is non-monotonic: noise is enhanced by increasing $\lambda_{\text{ex}}$ for small loops, but saturates at $\lambda_{\text{ex}}\sim 100\text{--}200$ kbp, and decreases after that. It is notable that this size is close to the median CTCF-cohesin loop size in mammals~\cite{Gassler2017,Brackley2017,Banigan2023}. The dependence on $\lambda_{\text{ex}}$ mirrors that on $d_{\text{TU}}$ (\fig\ref{fig:TU_noise}), and may be due to the fact that cohesin loops favor the formation of TU loops in their interior, but hinder interactions between TUs straddling the inside and outside of a cohesin loop~\cite{Brackley2021,Banigan2023}. 

To understand the reason underlying the enhancement of noise with $N_{\text{ex}}$, we hypothesize that this is driven by different cohesin loops formed through the stochastic 1D extrusion process in different simulations (or different cells). To test this explanation, we consider a population of chromatin fibers with the \textit{same} set of cohesin loops. Indeed, noise decreases rather than increases in this condition [orange curve in \fig\ref{fig:extrusion}(a)], as the loops compactify the chromatin fiber and stabilize clusters. These results imply that cohesin loops contribute more significantly to \textit{extrinsic} noise within a cell population than to intrinsic noise within a single cell. Similar to intrinsic noise, cohesin-mediated noise can also be understood by looking at the variability in TU clustering in 3D, whose phase diagram strongly correlates with that of noise [\fig\ref{fig:extrusion}(d)].

\paragraph{Conclusions.} In summary, we have studied a simple polymer model for chromatin to unveil the biophysical principles underpinning transcriptional noise. We find that intrinsic noise, determining transcriptional variability within a single cell, is intimately linked to the clustering of transcription units through bridging-induced phase separation~\cite{Brackley2013,Brackley2016}. Clustering arises from cooperative binding, which in turn induces temporal correlations which enhance noise; if it is disrupted (for instance by abrogating multivalent chromatin-protein binding), noise sharply decreases. Cohesin loops, emerging through ATP-mediated loop extrusion halted at CTCF~\cite{Fudenberg2016}, do not greatly affect average transcriptional activity, in line with experiments~\cite{Rao2017,Hsieh2022}, but contribute instead to extrinsic noise, or transcriptional heterogeneity within a cell population. This is because different CTCF-cohesin loops form stochastically via extrusion in different cells. As CTCF is, in evolutionary terms, a relatively recent addition to the repertoire of proteins responsible for chromatin organization, we speculate our finding that loop extrusion enhances noise may explain why evolutionarily young genes tend to be noisier than older ones~\cite{Capra2013}.

Intrinsic and extrinsic noise are both important and are arguably regulated in cells. The former controls transcriptional bursting~\cite{Fukaya2016,Tunnacliffe2020,Sukys2024}, key to fast transcriptional response (e.g.,~\cite{Ginley2024}). The latter is crucial to endow a cell population with the ability to react or differentiate in response to external cues, which is relevant during development and in disease. We suggest that the patterning of TUs and CTCF binding sites provide distinct genetic and epigenetic handles to regulate transcriptional noise. These predictions may be tested with targeted single-cell transcriptomics and single-molecule RNA and/or DNA FISH experiments.

\begin{acknowledgments}
We thank Y. S. Ng and A. Sukys for useful discussions. This work received support from the UK Medical Research Council (MC\_UU\_00007/13; MC\_UU\_00035/6) and the Wellcome Trust (223097/Z/21/Z).\vspace{5pt}
\end{acknowledgments}

M.C., C.B., and G.F. contributed equally to this work.

%

\pagebreak
\cleardoublepage 


\title{\papertitle: Supplemental Material}

\maketitle

\setcounter{equation}{0}
\renewcommand{\theequation}{S\arabic{equation}}

\setcounter{figure}{0}
\renewcommand{\thefigure}{S\arabic{figure}}

\setcounter{table}{0}
\renewcommand{\thetable}{S\arabic{table}}

\section{Simulation Methods}

In the following, we explain the molecular dynamics procedure used in this work for simulating the three-dimensional (3D) folding of the chromatin fiber and its transcriptional activity. Our simulation framework is in line with previous work that models chromatin at a coarse-grained level using a direct or ``bottom-up'' approach, focusing on the biophysical mechanisms driving chromatin folding~\cite{Brackley2020,Chiang2022TIG}.

\subsection{Coarse-Grained Modeling of the Chromatin Fiber}

We model the chromatin fiber as a semi-flexible bead-and-spring chain of $N$ beads, each representing $1$ kbp of chromatin and having a diameter $\sigma_b$. We define a subset of $N_{\text{TU}}$ beads as transcription units (TUs), modeling \textit{cis}-regulatory elements, which are spaced uniformly along the chain at separation $d_{\text{TU}}$ unless otherwise specified. For simulations where we vary $d_{\text{TU}}$, we keep $N_{\text{TU}}$ constant, and the total number of beads of the chromatin chain is $N = (N_{\text{TU}}+1)d_{\text{TU}}-1$ (i.e., we have a segment of non-TU chromatin beads on both ends of the fiber).

The chromatin chain is governed by several potentials. First, there is steric repulsion between beads, modeled via the Weeks-Chandler-Andersen potential
\begin{equation}
    U_{\text{WCA}}(r) = 4k_BT\left[\left(\frac{\sigma_b}{r}\right)^{12} - \left(\frac{\sigma_b}{r}\right)^6+\frac{1}{4}\right] \;,
\end{equation}
for $r < 2^{1/6}\sigma_b$ and $0$ otherwise, where $r$ is the separation between two beads, $k_B$ is the Boltzmann constant and $T$ the temperature of the system. Next, to ensure chain connectivity, consecutive beads along the chain are bonded via the finite extensible non-linear elastic (FENE) potential
\begin{equation}
    U_{\text{FENE}}(r)=-\frac{K_{F}R_0^2}{2}\log\left[1-\left(\frac{r}{R_0}\right)^2\right] \;,
\end{equation}
for $r < R_0$ and $\infty$ otherwise, where the spring constant $K_F = 30k_BT/\sigma_b^2$ and the maximum bond separation $R_0 = 1.6\sigma_b$. This, combined with the WCA potential, gives an equilibrium bond length of ${\sim}1\sigma_b$. To account for chromatin stiffness, we impose an angle cosine potential (i.e., a Kratky-Porod model)
\begin{equation}
    U_{\text{bend}}(\theta) = K_{\text{bend}}(1-\cos\theta) \;,
\end{equation}
where $\theta$ is the angle between three consecutive beads defined by $\cos(\theta) = \vec{t}_{i+1}\cdot\vec{t}_{i}$, with $\vec{t}_i = (\vec{r}_{i+1}-\vec{r}_{i})/\abs{\vec{r}_{i+1}-\vec{r}_{i}}$, and $K_{\text{bend}}$ controls the degree of bending and is related to the fiber's persistence length $\ell_p$ by $K_{\text{bend}} = k_BT\ell_p/\sigma_b$. We set $\ell_p = 4\sigma_b$, consistent with previous work modeling chromatin at this resolution~\cite{Buckle2018,Chiang2022}.

\begin{figure*}[t!]
  \centering
  \includegraphics[width=\textwidth]{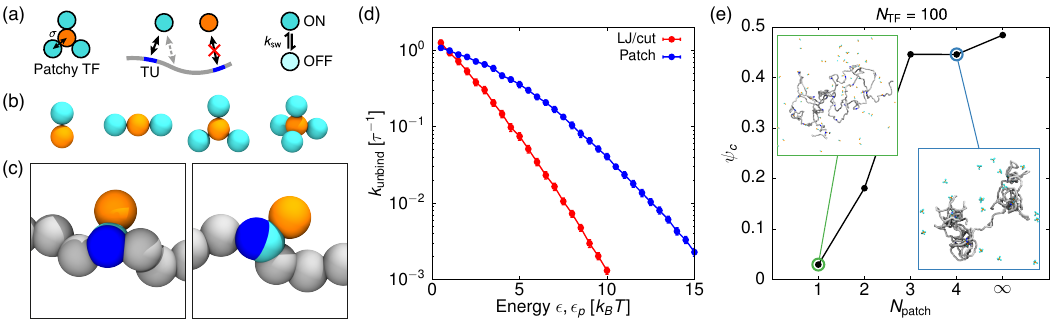}
  \caption{Modeling TFs as patchy rigid bodies to limit their valency in chromatin binding. (a) Schematics of the modified TFs. The cyan patches on a TF are attracted to TUs when the TF is switched on, and the orange core bead interacts repulsively with the TU. (b) The geometry and relative orientation of the chromatin-binding patches on a TF as the number of patches $N_{\text{patch}}$ increases from $1$ to $4$. (c) Simulation snapshots illustrating that each patch on a TF can typically bind to a single bead (\textit{left}) or two consecutive beads (\textit{right}). (d) Comparing the unbinding rates $k_{\text{unbind}}$ from chromatin for a non-patchy and a single-patch TF when varying the energy of the respective potentials (i.e., $U_{\text{LJ/cut}}$ for non-patchy and $U_{\text{patch}}$ for patchy TFs). (e) A plot showing the fraction of TFs which form clusters $\psi_c$ as a function of $N_{\text{patch}}$ (note that $N_{\text{patch}} = \infty$ corresponds to the non-patchy case). The line connecting the data points is drawn to guide the eye. Insets show representative snapshots of the system for the one-patch and four-patch cases. In all patchy-TF simulations we consider $d_{\text{TU}} = 30$ kbp and $N_{\text{TU}} = 30$.}
  \label{sfig:TF_patch}
\end{figure*}

\subsection{Transcription Factor Binding}

A crucial component driving the 3D folding of chromatin is its interaction with transcription factors (TFs). For simplicity, we model these factors as spherical beads with the same diameter $\sigma_b$ as chromatin beads; these can diffuse freely within the simulation box. While only interacting repulsively among themselves (via the WCA potential), TFs are multivalent in their interaction with chromatin, allowing them to bridge between two or more chromatin segments. In particular, we model this interaction using the truncated and shifted Lennard-Jones (LJ) potential
\begin{equation}
    U_{\text{LJ/cut}}(r) = \frac{1}{\mathcal{N}_{\text{LJ}}}\left[U_{\text{LJ}}(r)-U_{\text{LJ}}(r_c)\right] \;,
\end{equation}
for $r < r_c = 1.8\sigma_b$ and $0$ otherwise, where
\begin{equation}
    U_{\text{LJ}}(r) = 4\epsilon\left[\left(\frac{\sigma_b}{r}\right)^{12}-\left(\frac{\sigma_b}{r}\right)^{6}\right]
\end{equation}
is the standard LJ potential and 
\begin{equation}
    \mathcal{N}_{\text{LJ}} = 1 + 4\left[\left(\frac{\sigma_b}{r_c}\right)^{12}-\left(\frac{\sigma_b}{r_c}\right)^6\right]
\end{equation}
is a normalization factor that ensures the potential reaches $-\epsilon$ at its minimum. TFs bind strongly to chromatin TU beads, with $\epsilon = 7k_BT$, and non-specifically to other chromatin beads, with $\epsilon = 3k_BT$. To account for the dynamical turnover of constituents within TF clusters driven by bridging-induced phase separation (BIPS), TFs switch between a chromatin binding and non-binding state with rate $k_{\text{sw}} = 10^{-5}\tau^{-1}$, where $\tau$ is the time unit in the 3D simulations. When TFs are non-binding, they experience steric repulsion with all other beads, governed by the WCA potential.

\subsection{Modeling Transcription Factors\\ with Limited Valency}
 
To investigate how BIPS affects transcriptional noise, in a subset of simulations we modify the model such that TFs have limited valency in their interaction with chromatin beads. This is done by modeling TFs as patchy rigid bodies, as considered previously~\cite{Brackley2020JPhys}. Specifically, we attach extra beads, or patches, at a distance $\sigma_b$, to a core bead representing a TF [see \figs\ref{fig:TU_noise}(b) and \ref{sfig:TF_patch}(a)].

We consider patchy TFs with a valency from $N_{\text{patch}} = 1$ to $4$. Patches are positioned around the core such that they are equally spaced from one another to limit the superposition of the interactions of different patches. For instance, patches of a three-patch TF form an equilateral triangle, whereas those of a four-patch TF form a tetrahedryon. The precise geometry of the patches can be seen in \fig\ref{sfig:TF_patch}(b).  

The core bead of the TF now interacts repulsively with chromatin beads and the cores of other TFs, whereas each patch extending from the core is monovalent when the TF is switched on -- i.e., it binds to a single local chromatin segment and does not form bridges between non-local segments [\fig\ref{sfig:TF_patch}(c)]. In this way, the valency of the TF is determined by the number of patches it possesses. To enforce monovalency, we use the following potential for the patch-chromatin interaction:
\begin{equation}
  U_{\text{patch}} = \frac{\epsilon_p}{\mathcal{N}_p}\left(e^{-2ar}-2e^{-ar} - e^{-2ar_c} + 2e^{-ar_c}\right) \;,
\end{equation}
for $r < r_c = 1.2\sigma_b$ and $0$ otherwise, with $a = 1.75\sigma_b^{-1}$ and the normalization factor
\begin{equation}
    \mathcal{N}_p = 1 + \left(e^{-2ar_c}-2e^{-ar_c}\right)
\end{equation}
ensuring that the potential depth reaches $-\epsilon_p$. Previous work~\cite{Brackley2020JPhys} showed that using this potential, a patch usually binds to a single chromatin bead or sits between two consecutive beads but rarely bridges between non-consecutive beads. Note that patches only interact with chromatin beads and not with any other beads (i.e., zero energy for the interactions with other patches and with core TF beads), so that the effective size of a patchy TF remains similar to that of a non-patchy TF.  

To ensure a fair comparison in clustering dynamics, we choose $\epsilon_p$ such that the kinetics of binding to (or unbinding from) chromatin for these patchy TFs remain similar to that for non-patchy TFs. This calibration is important given that the shape and interaction range of the potentials $U_{\text{patch}}$ and $U_{\text{LJ/cut}}$ are different, and so using the same energy may give different binding and unbinding statistics for these two potentials. To this end, we measure the typical unbinding rate $k_{\text{unbind}}$ from chromatin for a non-patchy TF and a single-patch TF, both originally bound to a TU located at the midpoint of a short chromatin chain with $N = 100$ beads, as we vary $\epsilon$ and $\epsilon_p$. \fig\ref{sfig:TF_patch}(d) shows these rates, and from the plot we select $\epsilon_p$ values (for $U_{\text{patch}}$) that give comparable unbinding rates as those for $\epsilon$ values (for $U_{\text{LJ/cut}}$). In particular, we use $\epsilon_p = 5k_BT$ for non-specific binding (between TF patches and non-TU chromatin beads) and $\epsilon_p = 12 k_BT$ for specific interactions (between TF patches and TU beads). 

To explore how the TF's valency for chromatin binding affects BIPS, we vary the number of patches $N_{\text{patch}}$ on TFs and measure the fraction of TFs $\psi_c$ which form clusters [we say a TF is within a cluster if it is close to at least one other TF, based on the contact threshold $r_C = 3.5\sigma_b$; \fig\ref{sfig:TF_patch}(e)]. We find $\psi_c$ increases sharply with $N_{\text{patch}}$ and saturates when $N_{\text{patch}} \geq 3$, in line with our expectation that a higher valency enables more effective chromatin bridging and thus cluster formation.

\subsection{Loop Extrusion}

In some of our simulations, we incorporate chromatin loops mediated by structural maintenance of chromosome (SMC) complexes, such as cohesin, to study their effect on transcription. We model these loops by first running 1D loop extrusion (LE) simulations to generate an ensemble of loop anchor positions, and then imposing loops along the chromatin fiber according to these positions in the 3D polymer simulations. 

In the 1D extrusion simulations, we consider a total of $N_\text{ex}$ loop-extruding factors (or extruders). Following previous studies on LE dynamics~\cite{Fudenberg2016}, the two loop anchors associated with an extruder first bind to the chromatin fiber at beads $i$ and $i+1$, where $i$ is chosen randomly from $1$ to $N-1$ with uniform probability. They then move stochastically and divergently along the fiber (i.e., $i,i+1 \to i-1,i+2 \to i-2,i+3$, and so on) with velocity $v_{\text{ex}}$, and can unbind from the fiber with rate $k_{\text{off}}$. The ratio of these two parameters set the typical loop size, or processivity, given by $\lambda_{\text{ex}} = v_{\text{ex}}/k_{\text{off}}$. We vary $\lambda_{\text{ex}}$ in the simulations by changing $k_{\text{off}}$ and keeping $v_{\text{ex}}$ constant; in particular, we set $\tau_{\text{ex}} = 1~\text{bead}/v_{\text{ex}}$ as the unit of time in these 1D simulations. For simplicity, extruders that unbind from the fiber are re-attached to it at a different location within the same timestep. The two anchors of a loop extruder move independently of each other and halt if they collide with anchors of other extruders or reach the end of the chain. We run LE simulations for a duration of $10^4\tau_{\text{ex}}$ and use the final positions of the extruders' loop anchors for placing loops in the 3D polymer simulations, which is done by adding extra bonds between loop anchor beads via the potential
\begin{equation}
    U_{\text{ex}}(r) = U_{\text{WCA}}(r) + \frac{K_{\text{ex}}}{2}\left(r-r_0\right)^2 \;.
\end{equation}
Here, we set the bond stiffness $K_{\text{ex}} = 80 k_BT/\sigma_b^2$ and the bond length $r_0 = 1.5\sigma_b$. Note that we perform the 1D extrusion simulation for each run with a different random seed, so the positions of the loops are different between simulations, but their loop lengths are related and depend on the parameters $N_{\text{ex}}$ and $\lambda_{\text{ex}}$.

\subsection{Simulation Box and Integration Scheme}

We simulate the chromatin fiber in a periodic cubic box of length $L = 100\sigma_b$ using Langevin dynamics. Specifically, beads (both chromatin and TF) are governed by the equations of motion
\begin{equation}
    m\frac{\fd^2\vec{r}_i}{\fd t^2} = -\vec{\nabla}_i U - \gamma\frac{\fd\vec{r}_i}{\fd t} + \sqrt{2\gamma k_BT}\vec{\eta}_i(t) \;,
\end{equation}
where the first term on the right-hand side represents conservative forces derived from the total potential energy $U$ of the system, and the second term models damping due to a solvent (i.e., the nucleoplasm). The last term accounts for random forces due to collisions with solvent particles, with the noise vector $\vec{\eta}_i$ obeying the following averages
\begin{align}
  \avg{\eta_{i,\alpha}(t)} &= 0 \\
  \avg{\eta_{i,\alpha}(t)\eta_{j,\beta}(t')} &= \delta(t-t')\delta_{ij}\delta_{\alpha\beta} \;.
\end{align}
Here, $\delta(t-t')$ is the Dirac delta function, whereas $\delta_{ij}$ and $\delta_{\alpha\beta}$ are Kronecker deltas, with Latin indices running over beads and Greek indices over Cartesian components. For simplicity, we have assumed that all beads have the same mass $m$ and experience the same damping $\gamma$. The equations of motion are integrated using the velocity-Verlet scheme with timestep $\Delta t = 0.01\tau$, and this is performed using the Large-scale Atomic and Molecular Massively Parallel Simulator (LAMMPS) package~\cite{Plimpton1995}. 

\subsection{Initialization and Equilibration Procedure}

We initialize the chromatin chain as a random walk and run short simulations to allow it to relax into a self-avoiding chain. To avoid numerical divergence during this period, we initially model bonds between consecutive chromatin beads via the harmonic potential
\begin{equation}
  U_{\text{harm}} = \frac{K_h}{2}(r-r_0)^2 \;,
\end{equation}
with $K_h = 200 k_BT/\sigma_b^2$ and $r_0 = 1.1\sigma_b$, and we use a higher persistence length $\ell_p = 10\sigma_b$ to help remove any entanglements. Additionally, we use a soft potential for the pairwise repulsive interactions between beads, given by
\begin{equation}
  U_{\text{soft}}(r)= A \left[1+\cos\left(\frac{\pi r}{r_c}\right)\right] \;,
\end{equation}
for $r < r_c = 1.8\sigma_b$ and $0$ otherwise, and we set $A = 100 k_BT$ (interactions between TF and chromatin beads are purely repulsive in this period). We allow the system to equilibrate for a period of $10^2\tau$ using these potentials, before reverting to using FENE bonds and WCA potential for steric repulsion and lowering the persistence length to $\ell_p = 4\sigma_b$. The system is then evolved for further period of $5\times10^3\tau$.

For simulations incorporating extruded chromatin loops, we next add bonds associated with these loops incrementally to avoid introducing knots to the chromatin chain and generating numerical divergence. Specifically, for each loop, we first add a bond between the two beads located at the midpoint of the loop anchor beads. The bond is then moved divergently every $50\tau$ until reaching the desired anchor beads. Since loops can be nested, we load loops in the outmost layer first, then those in the second layer, and so on. We stop the simulation once all extrusion bonds have moved to their respective loop anchor beads. Note that this procedure avoids catenations and other entanglements.

We further equilibrate the system for a period of $5\times10^3\tau$ (for both the case with or without extruded loops), before switching on chromatin-TF interactions and equilibrating the fiber for $10^4\tau$. Finally, for the main production run, we perform the simulations for a period of $10^5\tau$, and we sample the conformation of the fiber and the transcriptional activity of each TU every $10^3\tau$, giving a total number of sampled time frames to be $M = 101$. We repeat the simulation for $N_{\text{sim}} = 100$ times (with different initial configurations and seeds for the random number generator) to sample variation in transcriptional activity and chromatin conformation.

\subsection{Mapping between Simulation and Physical Units}

For the 3D polymer simulations, we express distance in units of the bead size $\sigma_b$, time in units of $\tau$, and energy in units of $k_BT$. In a previous work~\cite{Buckle2018} with a similar model, we compared distances measured in simulations to those from FISH experiments, and we find a bead modeling $1$ kbp of chromatin has a diameter of $\sigma_b \sim 20~\text{nm}$. To map from simulation to real time, we note three important timescales in the simulations. First, there is the natural timescale $\tau = \sigma_b\sqrt{m/(k_BT)}$ that is used when performing the simulations. There is also the inertial time $\tau_{\text{in}} = m/\gamma$ and the Brownian time $\tau_B = \sigma_b^2/D$, where $D$ is the diffusion coefficient of a single bead. Large-scale conformational changes in the chromatin fiber typically occur over the scale of $\tau_B$; however, substituting realistic values for chromatin one finds $\tau_{\text{in}} \ll \tau \ll \tau_B$, and due to numerical stability each timestep $\Delta t$ has to be smaller than $\tau$. In order to reach the Brownian scale and avoid running unrealistically long simulations, we set $\gamma/m = 1$ (in simulation units), which is much smaller than in reality, such that $\tau_{\text{in}} = \tau = \tau_B$. In this way, we observe changes in chromatin structure in the simulations, while noting that dynamics over timescales smaller than $\tau_B$ is less realistic. To estimate the Brownian time, we note that $D$ is related to the damping constant $\gamma$ via the Einstein relation $D = k_BT/\gamma$, and so $\tau_B = 3\pi\nu\sigma_b^3/(k_BT)$, where $\nu$ the viscosity of the solvent and we assumed that Stoke's law holds for a spherical bead. Substituting $T = 300~\text{K}$ and considering the nucleoplasm has a viscosity of $\nu = 100~\text{cP}$, we find $\tau_B \sim 2~\text{ms}$. 

\section{Measuring Transcriptional\\ Activity and Clustering of Transcription Units}

As discussed briefly in the main text, we consider a TU to be undergoing transcription when a TF is associated with it, which we define to be when the TF is within a distance of $r_C = 3.5\sigma_b$ from the TU. More quantitatively, we define the transcription state of TU $i$ as $s_i \in \{+1,-1\}$, where $s_i = +1$ if the TU is being transcribed and $-1$ otherwise. This state variable can be expressed as
\begin{equation}
    s_i(t) = \frac{2n_i(t)}{\text{max}[n_i(t),1]}-1
\end{equation}
with
\begin{equation}
    n_i(t) = \sum_{j=1}^{N_{\text{TF}}}\Theta[r_C-r_{ij}(t)]  \;,
\end{equation}
where $\Theta(x)$ is the Heaviside step function, $N_{\text{TF}}$ is the number of TFs, and $r_{ij}$ is the separation between TU $i$ and TF $j$ (we use the position of the core bead for simulations where we model TFs as patchy rigid bodies). The fraction of time $\phi_i$ that TU $i$ is transcribed is then given by
\begin{equation}
    \phi_i = \frac{1}{2M}\sum_{n=0}^{M-1} \left[s_i(t_n)+1\right] \;,
\end{equation}
where $\{t_n\}$ are the time frames ($M$ in total) at which we sample the system in each simulation.

\begin{figure*}[ht!]
  \centering
  \begin{minipage}{\textwidth}
  \includegraphics{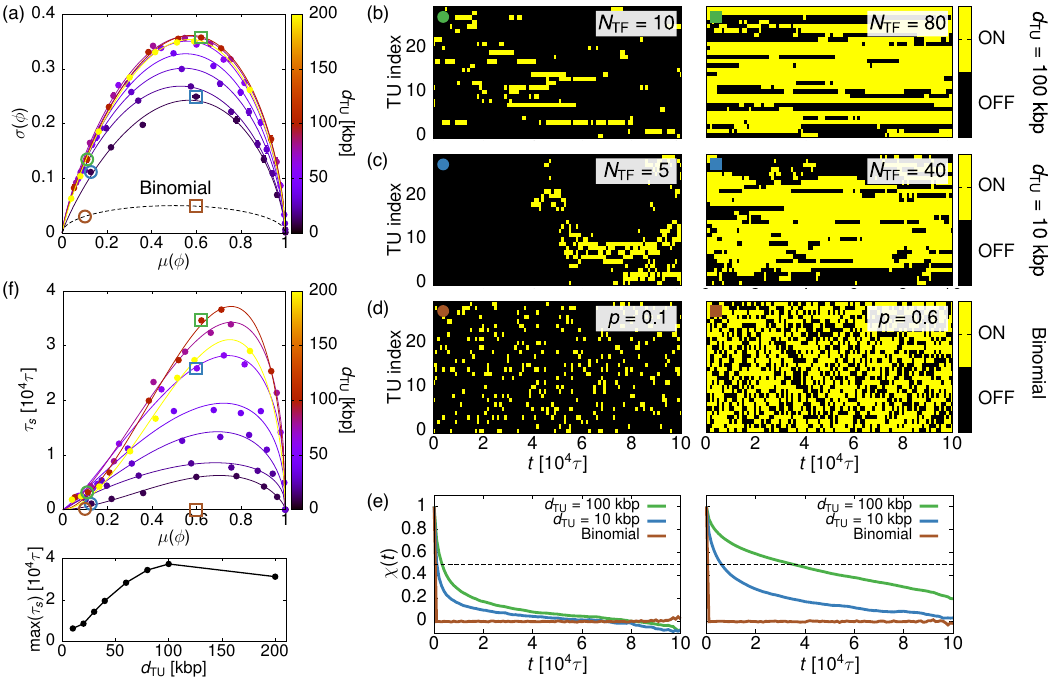}
  \caption{Temporal correlation in transcriptional activity of TUs. (a) Transcription boomerangs for different TU spacings $d_{\text{TU}}$ as shown in \fig\ref{fig:TU_noise}(a). (b)--(d) Kymographs showing the transcription status $s_i$ [i.e., ON ($+1$) or OFF ($-1$)] of each TU over time for cases with low activity $\mu \sim 0.1$ [points marked by circles in (a); \textit{left}] and intermediate activity $\mu \sim 0.6$ [points marked by squares; \textit{right}], for (b) $d_{\text{TU}} = 100$ kbp and (c) $d_{\text{TU}} = 10$ kbp, and for (d) the binomial model (i.e., a sequence of Bernoulli trials), where we consider the same as the number of sampling events ($M = 101$) as in the other cases (i.e., the time interval between frame/event is $10^3\tau$). (e) The autocorrelation $\chi(t) = \tilde{\chi}(t)/\tilde{\chi}(0)$ in transcription [see \eqn\eqref{eqn:auto_corr}] for the points considered in (b)--(d). The dashed line indicates the threshold ($\chi^* = 0.5$) we use for defining the correlation time $\tau_s$. (f) Measured $\tau_s$ as a function of $\mu$ for different $d_{\text{TU}}$ (\textit{top}). As with transcription boomerangs, we fit the curve $\tau_s(\mu) = \frac{A_{\tau_s}\mu^{\alpha}(1-\mu)^{\beta}}{\nu^{\alpha}(1-\nu)^{\beta}}$, with $\nu = \frac{\alpha}{\alpha+\beta}$, to guide the eye and extract the maximum correlation time [i.e., $\text{max}(\tau_s) = A_{\tau_s}$], which we plot as a function of $d_{\text{TU}}$ in the bottom panel. Error bars representing the standard error on the mean (from averaging over all TUs) are shown for both axes in all boomerangs, but are smaller than the data points.}
  \label{sfig:trans_kymo}
  \end{minipage}
  \begin{minipage}{\textwidth} 
  \includegraphics{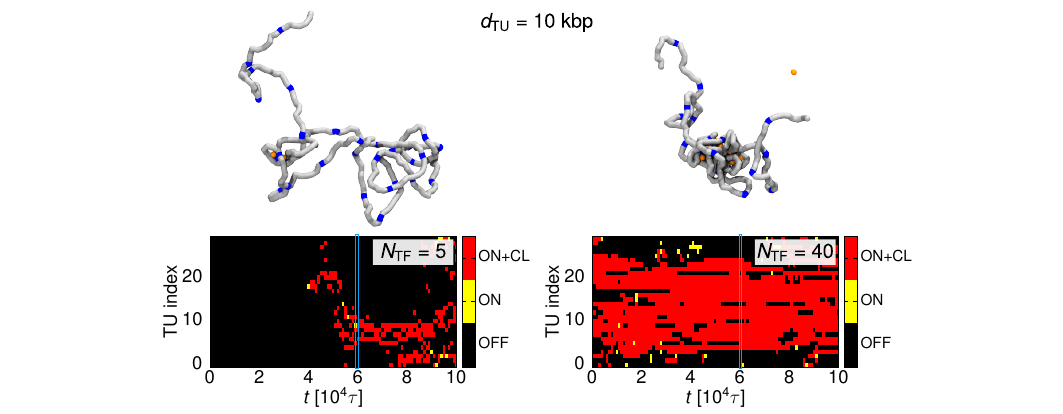}
  \caption{Transcriptional activity is linked to the clustering of TUs. Here, we replot kymographs shown in \fig\ref{sfig:trans_kymo}(c) with an additional state (red) for the case where TUs are transcribed and located within a cluster at the same time (ON+CL). This is also highlighted by snapshots above the kymographs showing the clustering of TUs at the time frame marked by the blue rectangle.}
  \label{sfig:trans_clust_kymo}
  \end{minipage}
\end{figure*}

\begin{figure*}[th!]
  \centering
  \begin{minipage}{\textwidth}
  \includegraphics{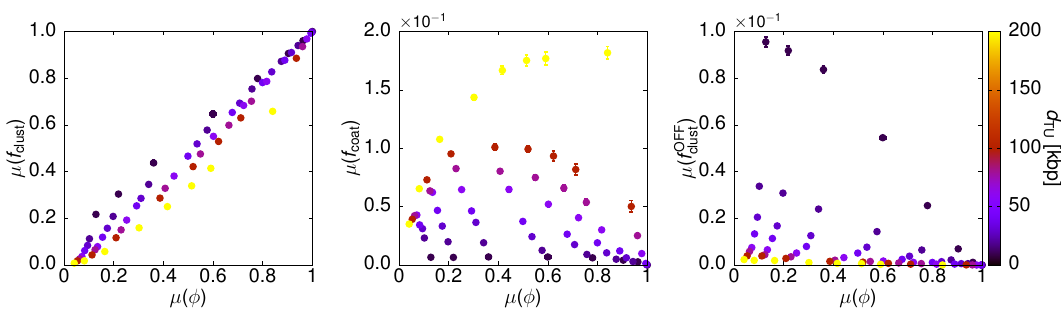}
  \caption{The contributions of TF coating and TU clustering mediated by BIPS to transcription. Here, we show scatterplots of the mean of $f_{\text{clust}}$, $f_{\text{coat}}$, and $f_{\text{clust}}^{\text{OFF}}$ as a function of transcriptional activity $\mu(\phi)$ for different TU spacings $d_{\text{TU}}$. Error bars representing the standard error on the mean (from averaging over all TUs) are shown for both axes for all plots but are smaller than the data points in some cases (same for \fig\ref{sfig:trans_clust_cov}).}
  \label{sfig:trans_clust_avg}
  \vspace{10pt}
  \end{minipage}
  \begin{minipage}{\textwidth}
  \includegraphics{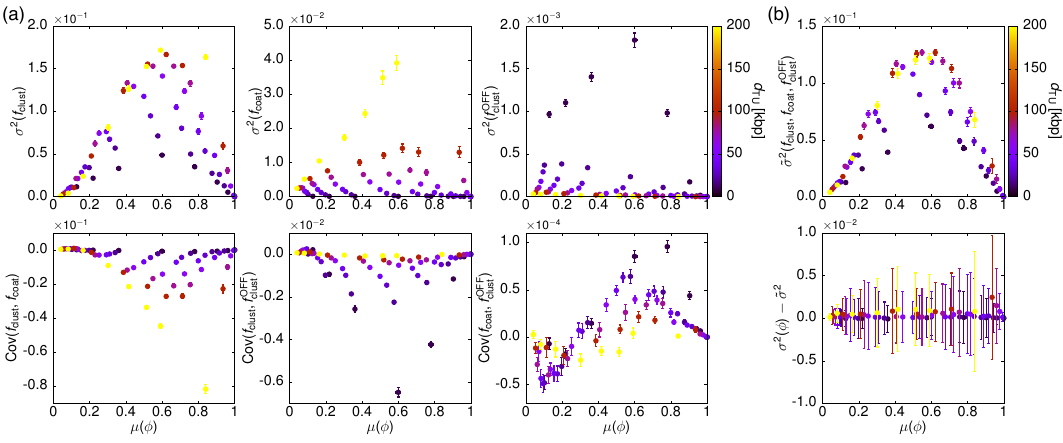}
  \caption{The contributions of TF coating and TU clustering mediated by BIPS to transcriptional noise. (a) Scatterplots showing the variance (\textit{top}) and covariances (\textit{bottom}) of $f_{\text{clust}}$, $f_{\text{coat}}$, and $f_{\text{clust}}^{\text{OFF}}$ as a function of transcriptional activity $\mu(\phi)$ for different TU spacings $d_{\text{TU}}$. (b) The resulting boomerangs from combining these fluctuation contributions (\textit{top}) and their deviation from the actual transcription boomerangs $\sigma^2(\phi)$ (\textit{bottom}).}
  \label{sfig:trans_clust_cov}
  \end{minipage}
\end{figure*}

To dissect the molecular mechanisms underpinning the overdispersion in transcriptional activity that cannot be captured by the binomial model, we plot kymographs of the transcription state $s_i$ of each TU [\figs\ref{sfig:trans_kymo}(a)--(d)]. One key feature observed is that TUs remain continuously transcribed (or not transcribed) for longer than expected compared to the binomial model. We quantify this by the autocorrelation function
\begin{equation}
  \tilde{\chi}(t) = \avg{s_i(t+t')s_i(t')}_{i,t'}-\avg{s_i(t')}_{i,t'}^2\;,
  \label{eqn:auto_corr}
\end{equation}
where the average is taken over time, TUs, and simulations, and we consider the normalized value $\chi(t) = \tilde{\chi}(t)/\tilde{\chi}(0)$ such that $\abs{\chi} \in [0,1]$. \fig\ref{sfig:trans_kymo}(e) shows that $\chi$ decays more slowly for $d_{\text{TU}} = 100$ kbp compared to $10$ kbp, as well as to the binomial model. We further define the correlation time $\tau_s$ as the time it takes for $\chi$ to decay to $1/2$. \fig\ref{sfig:trans_kymo}(f), plotting $\tau_s$ as a function of $\mu(\phi)$ for different $d_{\text{TU}}$, reveals that the correlation time increases with TU spacing, reaching a maximum at around $d_{\text{TU}}\sim 100$ kbp, before decreasing as TUs become further apart [see also the bottom panel of \fig\ref{sfig:trans_kymo}(f)]. This non-monotonic behavior is consistent with that for the transcriptional noise $\sigma(\phi)$, suggesting that noise is related to the degree of temporal correlation in transcription.

We reason the correlated dynamics in transcription is largely driven by the formation of TU clusters as a result of BIPS. In \fig\ref{sfig:trans_clust_kymo}, we replot the transcription kymographs shown in \fig\ref{sfig:trans_kymo}(c) but include an additional state (ON+CL; colored in red) for the case where TUs are both being transcribed and within clusters. Specifically, we say a TU to be within a cluster if it is within a contact threshold of $r_C = 3.5\sigma_b$ from another TU, and a cluster is defined to be a group containing at least two TUs. The kymographs show that in most cases transcription and TU clustering occur concomitantly. Clustering stabilizes the TU-TF interaction, enabling TU to be transcribed continuously for a longer period than by chance.

\pagebreak 

\subsection{Relating Clustering of Transcription Units\\ to Their Transcriptional Activity}

In \fig\ref{fig:TU_noise} we discuss the similarity between transcriptional noise $\sigma(\phi)$ and the fluctuation in the fraction of time a TU is within a cluster $\sigma(f_{\text{clust}})$. However, a closer inspection reveals that the trend of how $\sigma(\phi)$ varies with $d_{\text{TU}}$ is not fully recapitulated by that of $\sigma(f_{\text{clust}})$ [see bottom panels of \figs\ref{fig:TU_noise}(a) and (c)], suggesting that there are additional sources of fluctuations influencing noise. 

A more detailed analysis shows that there are several contributions to the fraction of time a TU is transcribed, which can be expressed as
\begin{equation}
  \phi = f_{\text{clust}} + f_{\text{coat}} - f_{\text{clust}}^{\text{OFF}} \;,
  \label{eqn:trans_comps}
\end{equation}
where $f_{\text{coat}}$ is the fraction of time a TF is coating on a TU (i.e., the TF is close to a TU but does not form a bridge between two or more TUs), and $f_{\text{clust}}^{\text{OFF}}$ is the fraction of time where the TU is located within a cluster but is not transcribed. \fig\ref{sfig:trans_clust_avg} shows how the average of these three contributions [i.e., $\mu(f_{\text{clust}})$, $\mu(f_{\text{coat}})$, and $\mu(f_{\text{clust}}^{\text{OFF}})$] vary as a function of the mean transcriptional activity $\mu(\phi)$. In particular, we see that $\mu(f_{\text{clust}})$ strongly correlates with $\mu(\phi)$, suggesting that the formation of clusters (i.e., transcription factories) plays a crucial role in transcription. Nevertheless, as the spacing between TUs increases, transcription driven by TF coating becomes more likely, as loops become harder to form. The case where TUs are in clusters and are not being transcribed is most likely to happen when the spacing between TU is small (${\leq}9.5\%$ of the time for $d_{\text{TU}} = 10$ kbp), and it becomes increasingly rare for larger spacings (${\leq}0.2\%$ of the time for $d_{\text{TU}} = 200$ kbp). To obtain an estimate of the variance of $\phi$ from \eqn\eqref{eqn:trans_comps}, we expand the fluctuation of $\phi$ around its mean, i.e., $\delta\phi = \phi-\mu(\phi)$, up to first order in terms of the observables $f_{\text{clust}}$, $f_{\text{coat}}$, and $f_{\text{clust}}^{\text{OFF}}$. Applying the definition of variance then allows us to write
\begin{widetext}
\begin{equation}
\begin{split}
  \sigma^2(\phi) \approx \tilde{\sigma}^2(f_{\text{clust}},f_{\text{coat}},f_{\text{clust}}^{\text{OFF}}) \equiv &\; \sigma^2(f_{\text{clust}}) + \sigma^2(f_{\text{coat}}) + \sigma^2(f_{\text{clust}}^{\text{OFF}}) \\
  &+ 2\,\text{Cov}\left(f_{\text{clust}},f_{\text{coat}}\right) - 2\,\text{Cov}\left(f_{\text{clust}},f_{\text{clust}}^{\text{OFF}}\right) - 2\,\text{Cov}\left(f_{\text{coat}},f_{\text{clust}}^{\text{OFF}}\right) \;,
\end{split}
\end{equation}
\end{widetext}
where $\text{Cov}(x,y)$ denotes the covariance of the observables $x$ and $y$. \fig\ref{sfig:trans_clust_cov}(a) shows the variances and covariances of $f_{\text{clust}}$, $f_{\text{coat}}$, and $f_{\text{clust}}^{\text{OFF}}$. Reassuringly, we see that by combining all these contributions one does recover the actual transcription boomerangs shown in the main text [\fig\ref{sfig:trans_clust_cov}(b)]. More importantly, we find $\sigma^2(f_{\text{clust}}) > \sigma^2(f_{\text{coat}}) > \sigma^2(f_{\text{clust}}^{\text{OFF}})$, which is in line with our expectation that the fluctuation in TU clustering plays the most dominant role in driving transcriptional noise. Additionally, we observe that the main contribution from the covariances is that of $f_{\text{clust}}$ and $f_{\text{coat}}$; since they are negatively correlated, particularly at large $d_{\text{TU}}$, this explains why the trends for $\sigma(\phi)$ and $\sigma(f_{\text{clust}})$ differ in the regime of large $d_{\text{TU}}$ [see \figs\ref{fig:TU_noise}(a) and (c)].

\begin{figure}[t!]
  \includegraphics[width=0.9\columnwidth]{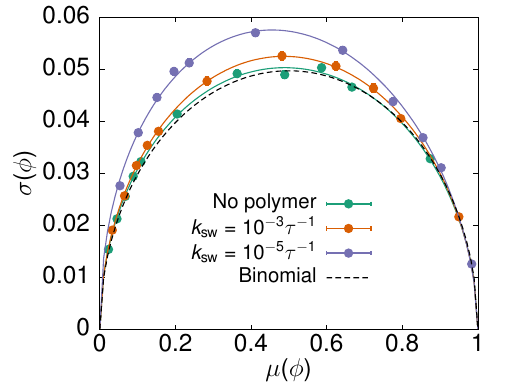}
  \caption{Transcription boomerangs from varying the switching rate $k_{\text{sw}}$ of TFs (orange and purple curves) and removing the polymeric connection between consecutive TUs (green curve). Here $d_{\text{TU}} = 30$ kbp and $N_{\text{TU}} = 30$, and the dashed boomerang corresponds to the binomial model.  We consider $N_{\text{TF}} = 20,40,60,80,100,200,400,600,800,1000,2000$ to construct the boomerangs. Error bars representing the standard error on the mean (from averaging over all TUs) are shown for both axes in all boomerangs but are smaller than the data points (same for some of subsequent boomerang plots).}
  \label{sfig:vary_switch}
\end{figure}

\begin{figure}[t!]
  \centering
  \includegraphics[width=0.9\columnwidth]{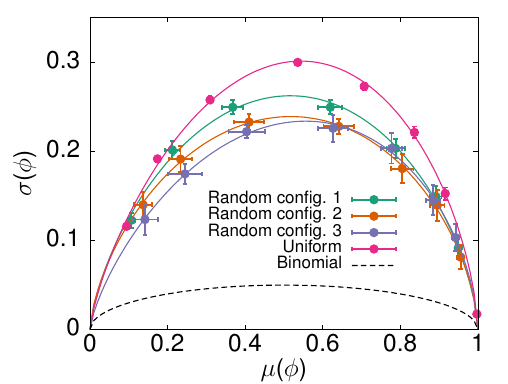}
  \caption{Transcription boomerangs for three different random configurations of TU positioning, keeping the global linear density of TUs the same as for the uniform case (shown here in pink, with $d_{\text{TU}} = 30$ kbp). We use $N_{\text{TU}} = 30$, and the dashed boomerang corresponds to the binomial model. We consider $N_{\text{TF}} = 5,10,20,40,60,80,100$ to construct the boomerangs.}
  \label{sfig:TU_random}
\end{figure}

\section{Other Simulation Setups and Their Effect on Transcriptional Noise}

We have also considered other simulation setups to further understand the mechanisms determining the relation between mean expression and transcriptional noise. For instance, when investigating how noise varies with the number of chromatin-binding patches on TFs, we find that having a single patch is still insufficient to fully recover the binomial model, with residual overdispersion [\fig\ref{fig:TU_noise}(b)]. This points to other biophysical mechanisms determining noise. Since the ability to form TU clusters (as a result of BIPS) is strongly implicated in noise, we hypothesize that other processes which weaken cluster formation will also reduce noise. One such mechanism is to increase the rate $k_{\text{sw}}$ at which TFs switch between a chromatin-binding and non-binding state, as this reduces the effectiveness of TFs forming chromatin bridges~\cite{Brackley2017biophysj}. Indeed, as shown in \fig\ref{sfig:vary_switch} (orange curve), a higher $k_{\text{sw}}$ lowers the level of overdispersion. Nevertheless, we still find the transcription boomerang to be higher than that for the binomial model. 

As noise is related to transcriptional and structural correlations, we envisage the fact that TUs are chained together as a polymer also provides correlations in their dynamics that can contribute to noise. To validate this, we remove the polymeric segments linking between consecutive TUs and simulate them as freely diffusive beads (we also keep the higher TF switching rate $k_{\text{sw}}$). Reassuringly, in this condition, the transcription boomerang (\fig\ref{sfig:vary_switch}, green curve) recovers that for the binomial baseline (dotted curve).

Another simulation setup discussed in the main text involves placing the TUs randomly along the fiber (while keeping the overall density of TUs along the fiber the same as the uniform case), which is a setup closer to that of a realistic chromosome segment. \fig\ref{sfig:TU_random} shows the transcription boomerangs for three different random configurations of TU positions along the chromatin fiber, and in all of these configurations we find the level of noise is lower than that for the case where TUs are uniformly spaced. This is because the non-uniform spacing makes TUs that are closer together in 1D along the fiber form more stable 3D loops, while TUs further apart are less likely to interact. This reduces the variability in looping or clustering patterns and thus leads to a lower level of transcriptional noise. \vspace{15pt}

\section{Simulation Parameters}

In this section, we summarize the key parameter values used for the various simulations conducted in this work. For the transcription boomerangs shown in \figs\ref{fig:TU_noise}(a) and (c), we run simulations with non-patchy TFs and without loop extrusion, and we set the number of TUs to be $N_{\text{TU}} = 30$. The spacing $d_{\text{TU}}$ between TUs and the number of TFs $N_{\text{TF}}$ are varied as follows:

\begin{table}[H]
  \centering
  \begin{tabularx}{\columnwidth}{|c|X|} \hline
       \textbf{Parameter} & \textbf{Values} \\ \hline
       $d_{\text{TU}}$ & $10,20,30,40,60,80,100,200$ kbp \\ \hline
       $N_{\text{TF}}$ & $5,10,20,40,60,80,100$ \\ \hline
  \end{tabularx}
  \caption{Parameter values for the simulations with non-patchy TFs and without loop extrusion.}
\end{table}

For simulations with patchy TFs and without loop extrusion [\fig\ref{fig:TU_noise}(b)], we set $N_{\text{TU}} = 30$ and $d_{\text{TU}} = 30$ kbp. $N_{\text{TF}}$ is varied based on the number of patches $N_{\text{patches}}$ on the TF as follows:

\begin{table}[H]
  \centering
  \begin{tabularx}{\columnwidth}{|c|X|} \hline
       \textbf{$N_{\text{patches}}$} & \textbf{$N_{\text{TF}}$}\\ \hline
       $1$ & $20,40,60,80,100,200,400,600,800,1000,2000$ \\ \hline
       $2$ & $10,20,40,60,80,100,200$ \\ \hline
       $3$ & $10,20,40,60,80,100,200$ \\ \hline
       $4$ & $5,10,20,40,60,80,100$ \\ \hline
  \end{tabularx}
  \caption{Parameter values for the simulations with patchy TFs and without loop extrusion.}
\end{table}

To construct the phase diagrams shown in \figs\ref{fig:extrusion}(b) and (c), we run simulations with loop extrusion (and with non-patchy TFs) where we set $N_{\text{TU}} = 40$, $d_{\text{TU}} = 30$ kbp, and $N_{\text{TF}} = 50$. The number of loop-extruding factors $N_{\text{ex}}$ and their processivity $\lambda_{\text{ex}}$ are varied as follows:
\begin{table}[h]
  \centering
  \begin{tabularx}{\columnwidth}{|c|X|} \hline
     \textbf{Parameter} & \textbf{Values} \\ \hline
     $N_{\text{ex}}$ & $5,10,20,40,60,80,100$ \\ \hline
     $\lambda_{\text{ex}}$ & $10,30,50,100,150,200,250,300$ kbp \\ \hline
  \end{tabularx}
  \caption{Parameter values for the simulations with non-patchy TFs and with loop extrusion.}
\end{table}

\end{document}